\documentclass[11pt]{article}
\usepackage{epsf,a4wide,amssymb,cite}
\makeatletter

\@addtoreset{equation}{section}
\def\section{\@startsection {section}{1}{\z@}{-3.5ex plus -1ex minus
 -.2ex}{2.3ex plus .2ex}{\large\bf\centering}}
\def\section{\@startsection {section}{1}{\z@}{-0.5ex plus -1ex minus
 -.2ex}{0.5ex plus .2ex}{\bf}}
\def\subsection{\@startsection{subsection}{2}{\z@}{-3.25ex plus%
 -1ex minus -.2ex}{1.5ex plus .2ex}{\bf}}
\def\subsection{\@startsection{subsection}{2}{\z@}{-0.25ex plus%
 -1ex minus -.2ex}{0.5ex plus .2ex}{\bf}}
\def\subsubsection{\@startsection{subsubsection}{3}{\z@}{-3.25ex plus%
 -1ex minus -.2ex}{1.5ex plus .2ex}{\sl}}
\makeatother

\newcommand{\ato}{{\ensuremath{ a_2^{(1)} }}}
\newcommand{\att}{{\ensuremath{ a_2^{(2)} }}}

\newcommand{\be}{\begin{equation}}
\newcommand{\ee}{\end{equation}}
\newcommand{\bea}{\begin{eqnarray}}
\newcommand{\eea}{\end{eqnarray}}
\newcommand{\nn}{\nonumber\\}

\def\d{{\rm d}}
\def\ds{\displaystyle}

\def\kS #1,#2,#3,#4,#5,#6.{{\ensuremath{
  S_{#1 #2}\left( #3 {#6 \atop #4} #5 \right)}}}

\def\R #1,#2,#3,#4.{{\ensuremath{
  R\left( #1 {#4 \atop #2} #3 \right)}}}

\def\lb#1{{\langle #1 \rangle}}

\newcommand{\Rm}{{$R$-matrix}}
\newcommand{\Rms}{{$R$-matrices}}

\newcommand{\sm}{{$S$-matrix}}
\newcommand{\sms}{{$S$-matrices}}

\newcommand{\tm}{{transfer matrix}}
\newcommand{\tms}{{transfer matrices}}

\def\barA{{\overline A}}
\def\barB{{\overline B}}
\def\blank#1{}

\def\bP{{\bf P}}

\def\cdd{{\cdot}}
\def\cev#1.{{\langle{#1}|}}
\def\cP{{\check P}}
\def\cR{{\check R}}

\def\cS{{\check S}}

\def\e{{\epsilon}}

\def\gup{{GU$'$}}
\def\gup{{RU}}

\def\tR{{\tilde R}}

\def\sab{{\ensuremath{{\it s}^{}_{\!AB}}}}

\def\vec#1.{{|{#1}\rangle}}

\def\cev#1.{{\langle{#1}|}}

\newcommand{\om}{\omega}

\def\vec#1.{{ \left| #1 \right\rangle}}

\begin{document}
\parindent 12pt
\parskip 9pt

{
\parskip 0pt
\newpage
\begin{titlepage}
\begin{flushright}
DFUB-98-18\\
KCL-MTH-98-38\\
hep-th/9810006\\
% version 1.6
September 30 1998\\[3cm]
\end{flushright}
\begin{center}
{\Large{\bf
Non-unitarity in quantum affine Toda theory\\
and perturbed conformal field theory
}}\\[1.4cm]
{\large G\'abor Tak\'acs%
\footnote{e-mail: takacs@bo.infn.it}%
$\;$ and $\;$ G\'erard Watts%
\footnote{e-mail: gmtw@mth.kcl.ac.uk}%
}\\[8mm]
${}^1$
{\em INFN -- Sezione di Bologna and Dip. di Fisica  -- Universit\'a di Bologna}
\\
{\em Via Irnerio 46, 40126 Bologna, Italy}\\[5mm]
${}^{2}$
{\em Department of Mathematics, King's College London,}\\
{\em Strand, London, WC2R 2LS, U.K.}\\[5mm]

{\bf{ABSTRACT}}
\end{center}
\begin{quote}
There has been some debate about the validity of quantum affine Toda
field theory at imaginary coupling, owing to the non-unitarity of the
action, and consequently of its usefulness as a model of perturbed
conformal field theory. Drawing on our recent work, we investigate
the two simplest affine Toda theories for which this is an issue --
\ato\ and \att.
By investigating the \sms\ of these theories before RSOS restriction,
we show that quantum Toda theory, (with or without RSOS restriction),
indeed  has some fundamental problems, but that these problems are of
two different sorts. For \ato, the scattering of solitons and
breathers is flawed in both classical and quantum theories, and RSOS
restriction cannot solve this problem. For \att\ however, while there
are no problems with breather--soliton scattering there are instead
difficulties with soliton-excited soliton scattering in the
unrestricted theory. After RSOS restriction, the problems with
kink--excited kink may be cured or may remain, depending in part on
the choice of gradation, as we found in \cite{us}. We comment on the
importance of regradations, and also on the survival of \Rm\ unitarity
and the \sm\ bootstrap in these circumstances.
\end{quote}
\vfill
\end{titlepage}
}

\section{Introduction}
\label{sec:one}
\setcounter{footnote}{0}

Affine Toda field theory has proved to be an extremely useful
description of a wide class of two dimensional scattering theories.
The main impetus behind the quantum treatments of these models has
been the description of massive scattering theories as perturbed
conformal field theories \cite{Zamo2}, and the fact that when treated
in the free-field formulation the actions of such perturbed conformal
field theories have the form of affine Toda theory actions
\cite{HMan2}.

These theories depend on a coupling constant $\beta$, and for real
values of $\beta$ describe a set of particles with purely elastic
\sms\ \cite{pests}. For imaginary values of $\beta$, the picture is
more involved.
The action is not real for real values of the fields, and typically
when one finds a reality condition on the fields which will leave the
action real, this will have a potential unbounded below. Clearly this
makes the quantum theory problematic.

However, in the classical theory the potential has discrete
vacua at real values of the fields, and there are solitonic solutions
interpolating these vacua and breather solutions of zero topological
charge; despite not taking real values, they have real  energy and
momentum \cite{OTUn2}, and indeed all the conserved quantities take
real values \cite{Free1}.
Given these properties of the solitons and breathers, it is not
surprising that much effort has been spent trying to quantise these
systems in spite of the other unpleasant properties.
One suggestion that has been made is that one simply needs to restrict
to some well--chosen subspace of well-behaved classical solutions.

There have been several attempts \cite{Holl1,NSL} to follow this scheme
and attempt a  semi-classical quantisation of the soliton and breather
sectors in the manner of \cite{DHN}. This has not been entirely
successful. While correctly `predicting' the quantum mass corrections
in some theories \cite{Holl2}, it has failed in some other cases
\cite{NSL} -- but neither of these papers actually applied the method
properly. The semi-classical calculation of mass corrections consists
of finding the spectrum of small oscillations around a background
soliton, but since the Hamiltonian cannot be shown to be Hermitian,
there is no guarantee that the results of \cite{Holl2,NSL} will have
any relation to the `real' answer%
\footnote{There is a wide literature on the spectra of non-self-adjoint
operators -- for an instructive example see e.g. \cite{EBDavies} and
the references therein.}.

Other problems have been highlighted by Khastgir and Sasaki
\cite{KSas1} who have found classical solutions with complex energy
which become singular in finite time, and solutions with negative real
energy, which would lead to problems if the desired quantum vacuum
could tunnel to these lower energy states, creating particles in the
process. However one might be able to argue that these are not in the
`proper' class of initial `breather and soliton' configurations.

The second route to quantise Toda theories has been through the use of
their quantum group symmetry. This has been carried out (at least in
part) for a number of Toda theories and has been successful in some
ways --  although this quantum--group procedure cannot be guaranteed
to give the correct \sms\ for the quantum solitons, the fact that
in every case so far investigated
\cite{smir,efthimiou,Holl1,smats,smats2,taka2} the \sm\ of the lowest
mass zero-topological charge bound states, or breathers, is identical
(after analytic continuation in  the coupling constant) to the
conjectured \sm\ of the lightest  mass particles in the real coupling
Toda theory argues very strongly that they are right.

Apart from Sine-Gordon theory ($a_1^{(1)}$), it is not expected that
the \sms\ for the fundamental solitons which result from the quantum
group method are unitary matrices, and the only known way to correct
this is quantum group or RSOS restriction \cite{smirresh,smir}.

However, in our recent paper \cite{us} on $\Phi_{1,5}$
perturbations of minimal models, we found examples of RSOS
restricted affine Toda field theories with `unitary' \sms\ for
the fundamental particle but nevertheless with a complex finite-size
spectrum.

This result motivated us to reexamine these issues in the two simplest
affine Toda field theories, $a_2^{(1)}$ and $a_2^{(2)}$.
These two theories have already been studied a lot; for $a_2^{(1)}$
there are papers on their classical and quantum integrability,
classical soliton solutions, classical breathers, semi-classical mass
corrections, quantum \sms, \sm\ bootstrap and RSOS
restrictions; for $a_2^{(2)}$ there is almost an equal amount of work done.

One of the objects of this paper is to highlight problems that can
already be found in the results in the literature but which have been
overlooked to date.
As we argue in section \ref{sec:2.1}, it is more important
that the eigenvalues of the \sms\ and higher \tms\ are
phases  than that the matrix is itself unitary, since this leads to a
real finite-size spectrum (to leading order); unfortunately  these
eigenvalues are not preserved under
RSOS restriction, where the relevant matrices are the
\tms\ of \cite{Zamo10,KFV}. Furthermore, the choice of
`gradation' is also important; while this leaves the eigenvalues of
the unrestricted \tms\ invariant, it has an effect on the form of the
\sm\ bootstrap and is an essential factor in the possible RSOS
restrictions, and hence in the eigenvalues of the restricted \tms.

We first investigate the unrestricted \sms\ of \ato\ and find that
while the fundamental soliton--soliton \sm\ has phase eigenvalues,
both the three-particle soliton \tm\ and the
the fundamental soliton--antisoliton \tm\ do not have phase
eigenvalues for generic coupling. While this problem may or may not be
cured by any particular RSOS restriction, we then show that the
soliton--breather scattering in
\ato\ also has intrinsic problems in both the quantum (section
\ref{sec:a21quantum}) and classical theory (section
\ref{a21classical}) which cannot be removed by RSOS restriction, and
that any theory containing both solitons and breathers will have a
complex spectrum.

For \att\ we find rather different results, in that the
unrestricted fundamental soliton--soliton scattering is well behaved,
but that the \sm\ of the fundamental soliton with the first excited
soliton is badly behaved.
For \att\ however, the classical theory (section
\ref{sec:a22classical}) does not seem to  reflect these difficulties.

First we start in section \ref{generalities} with some generalities on
\sms, their properties and constructions, and treat
\ato\ and \att\ in turn in sections
\ref{sec:a21quantum}--\ref{sec:a22classical}. In
section~\ref{sec:m314} we discuss a particular example which shows how
RSOS restriction  can fail in rendering the scattering theory
consistent and finish with our conclusions in
section~\ref{conclusions}.

%%%%%%%%%%%%%%%%% section 1 %%%%%%%%%%%%%%%%%%
\section{Generalities}
\label{generalities}

\subsection{\sms\ and unitarity}
\label{sec:2.1}

A quantum field theory is unitary if the inner product on the space of
states is positive definite and the \sm\ is unitary with respect to
this inner product. However, in statistical field theory this is too
strong a requirement, and a more natural one is that the spectrum is
real.

We will consider integrable field theories which have factorised \sms.
We consider two sorts of theories: those where the particles --
solitons and breathers -- are classified by their topological charge;
and later another where it is instead the vacua interpolated by the
particles which are labelled.

\subsubsection{\sms\ of particle theories}

The particles are divided into different `species', all the particles
of the same species having the same non-zero mass, and each species
having one or more `flavours'. We denote a particle of species $A$,
flavour $i$ and rapidity $\theta$ by $A_i(\theta)$.

If the in state is a 2-particle state
$\vec \,A_i(\theta)\,B_j(\phi)\, ._{in}$ with $\theta>\phi$, then the
\sm\ is
\be
  \vec \,A_i(\theta)\,B_j(\phi)\, ._{in}
= S_{AB}(\theta-\phi)_{ij}^{kl}\,
  \vec \,A_k(\theta)\,B_l(\phi)\,._{out}
\;,
\ee
where $\theta-\phi$ is real and positive for physical scattering.
$S_{AB}$ is a unitary {\em matrix} if
\be
  \sum_{k,l}  S_{AB}(\theta)_{ij}^{kl}\,
         \left(\, S_{AB}(\theta)_{i'j'}^{kl}\,\right)^*
= \delta_{ii'}\delta_{jj'}
\;,
\label{eq:usab0}
\ee
or in matrix notation
\be
  S_{AB}(\theta)\, S_{AB}(\theta)^\dagger = 1
\;.
\ee
The \sm\ of a unitary field theory possesses several properties,
such as real analyticity (A), and may possess other discrete symmetries
such as parity (P), time reversal (T) and
charge-conjugation (C). These are given in \cite{Zamo7} as
{
\renewcommand{\arraystretch}{1.3}
\be
\begin{array}{c|cccc}
  & A & C & P & T \\
\hline\\[-5mm]
  S_{AB}(\theta)_{ij}^{kl}
= \;\;
&
\left( S_{AB}(-\theta)_{ij}^{kl} \right)^*
&
S_{\barA\,\barB}(\theta)_{\bar i\bar j}^{\bar k\bar l}
&
S_{BA}(\theta)_{ji}^{lk}
&
S_{BA}(\theta)_{lk}^{ji}
\end{array}
\ee
}
Using (A) and (PT) it is possible to rewrite (\ref{eq:usab0}) as
\be
  \sum_{k,l}  S_{AB}(\theta)_{ij}^{kl}\,
              S_{AB}(-\theta)_{kl}^{mn} = \delta_i^m\delta_j^n
\;,
\label{eq:usab1}
\ee
for all values of $\theta$,
which is often simply written in matrix notation
\be
S_{AB}(\theta) S_{AB}(-\theta) = 1
\;.
\label{eq:gu}
\ee
This is usually known as `unitarity', but the only Toda theory for
which this is true seems to be Sine-Gordon theory, and it has no
direct relation to either unitarity (\ref{eq:usab0}) nor to the
`\Rm\ unitarity' of the underlying quantum group structure.

An alternative notation for the \sm\ uses the matrix $\cS_{AB}$
which is simply related to $S_{AB}$ by
\be
  \cS_{AB}(\theta)_{ij}^{kl} = S_{AB}(\theta)_{ij}^{lk}
\;,
\label{eq:scs}
\ee
If $S$ satisfies (A) and (T) we can express unitarity (\ref{eq:usab0})
as the matrix equation
\be
  \cS_{AB}(\theta) \, \cS_{BA}(-\theta) = 1
\;,
\label{eq:gu2}
\ee
which is again also known as `unitarity'.
Since this is indeed the equation satisfied by the \sms\ derived by
the quantum group construction, we shall refer to this as
\Rm\ unitarity, or \gup\ for short.

Note that \gup\ is equivalent to (\ref{eq:gu}) if the \sm\ is (P)
invariant, which is the case in Sine-Gordon theory.

\subsubsection{Reality of the spectrum}

When we try to describe a two-dimensional statistical model in terms
of an effective scattering theory, it is not necessary for the space
of states to have a positive definite inner product, and indeed many
interesting and physically relevant models are of this sort, for
example the scaling Yang--Lee model \cite{CMus3}.
However, whether or not the \sm\ satisfies unitarity (\ref{eq:usab0})
for a particular choice of the Hilbert space inner product,
it is certainly necessary that the eigenvalues of the \sm\
are phases for real rapidities if the spectrum is to be real.
While we do not know the {\em exact} equations for the finite-size
spectrum in the general case%
\footnote{Note exact equations for the finite size spectrum are now
known for several non-trivial interacting field theories, see e.g.\
\cite{ExcitedStates}.}%
, we know that the leading order finite-size effects are given by
the Bethe Ansatz.

The Bethe Ansatz equations express the condition that the
wave-function of a multi-particle state on a circle is single-valued
under the operation of taking a particle around the circle.
For  a state of two particles $A_a(\theta_A)$ and $B_b(\theta_B)$ on a
circle of circumference $R$,
the wavefunctions $\psi_{ab}(\theta_A,\theta_B)$ must satisfy the two
simultaneous equations
\be
\begin{array}{l}
   S_{AB}(\theta_A - \theta_B)_{a\,b}^{a'b'}\, \psi_{a'b'}
\,
= \eta\, \exp(-i R m_A \sinh\theta_A)\,\psi_{ab}
\;,
\\[2mm]
   S_{BA}(\theta_B - \theta_A)_{b\,a}^{b'a'}\, \psi_{a'b'}
\,
=  (1/\eta)\,\exp(-i R m_B \sinh\theta_B) \, \psi_{ab}
\;,
\end{array}
\label{eq:bae0}
\ee
where $\eta$ is a phase depending on the relative statistics of
particles $A$ and $B$.
These equations are only consistent if the two matrices commute, but
in our situations they obey the stronger condition RU and so
consistency is guaranteed.
Diagonalising these equations, they reduce to
\be
\begin{array}{l}
   \sab(\theta_A - \theta_B)
\, \exp(i R m_A \sinh\theta_A)
= \eta
\;,\;\;\;\;
 \exp(i R ( m_A \sinh\theta_A + m_B \sinh\theta_B))
= 1
\;,
\end{array}
\label{eq:bae}
\ee
where \sab\ is an eigenvalues of the \sm\
$S_{AB}(\theta)$. For this equation to be satisfied with real
rapidities,
$\sab(\theta_A - \theta_B)$ should be a pure phase for real
$\theta_A,\theta_B$; if it is not a pure phase, there can be no
solutions with real rapidities, and consequently the two-particle
states satisfying this quantisation condition will have complex energy
and momentum.

The generalisation of (\ref{eq:bae}) to multi-particle states
of particles $A,B\ldots N$ of masses $m_A,\ldots$ and rapidities
$\theta_A\ldots$ is
\be
\begin{array}{l}
  t_A(\theta_A\,\ldots,\theta_N)
\, \exp(i R m_A \sinh\theta_A)
= \eta_{AB}\,\eta_{AC}\,\ldots\,\eta_{AN}
\;,
\\[1mm]
  t_B(\theta_A\,\ldots,\theta_N)
\, \exp(i R m_B \sinh\theta_B)
= \eta_{BC}\,\eta_{BD}\,\ldots\,\eta_{BA}
\;, \;\; \hbox{etc}
\end{array}
\label{eq:bae2}
\ee
where $\eta_{AB}$ are phases depending on the relative statistics of
the particles,  and $t_A$ , $t_B$ etc are simultaneous
eigenvalues of the commuting%
\footnote{
The \tms\ $T_A$, $T_B$ etc commute by virtue of the
Yang-Baxter relation for the \sms.} \tms\ $T_A, T_B$ etc
which are shown graphically in figure \ref{fig:tm}
{\begin{figure}[ht]
\[
\begin{array}{cc}
\epsfxsize=5cm
\epsfbox[0 0 220 135]{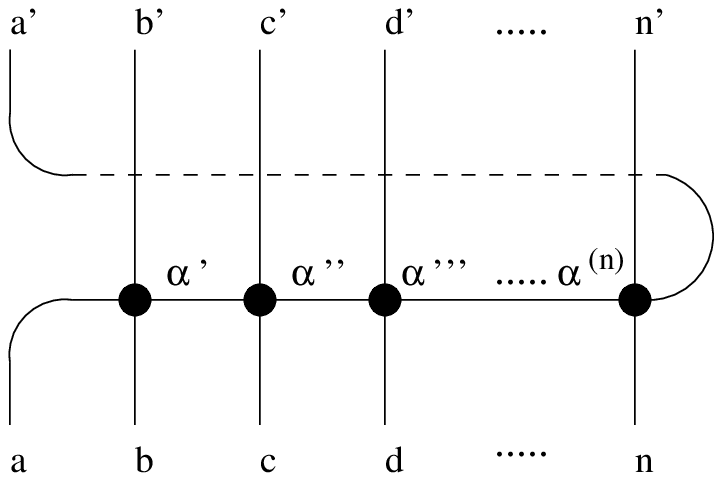}
&
\qquad\qquad
\epsfxsize=5cm
\epsfbox[0 0 220 135]{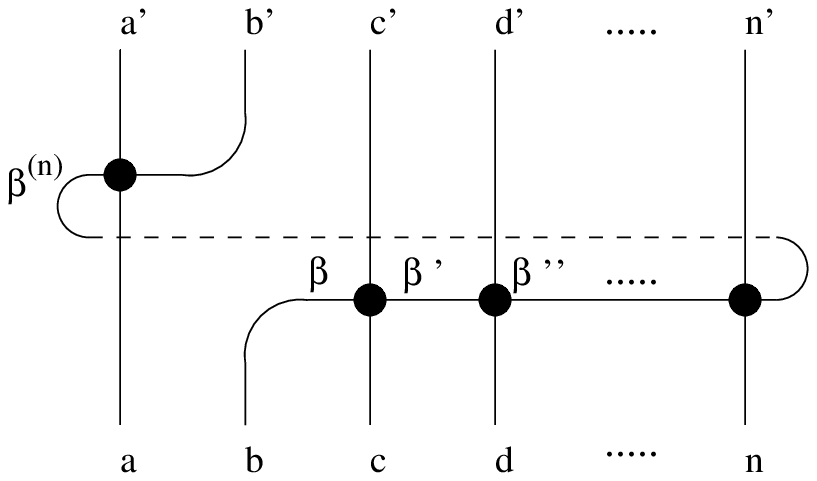}
\\[2mm]
  T_A(\theta_A,\ldots,\theta_N)_{a\,\,b\,\,c\,\,\ldots\,n}^{a'b'c'\ldots\,n'}
&\qquad\qquad
  T_B(\theta_A,\ldots,\theta_N)_{a\,\,b\,\,c\,\,\ldots\,n}^{a'b'c'\ldots\,n'}
\end{array}
\]
\caption{\small
The \tms\ $T_A$ and $T_B$.}
\label{fig:tm}
\end{figure}}%
and which are given by
\be
\begin{array}{l}
  T_A(\theta_A,\ldots,\theta_N)_{a\,\,b\,\,c\,\,\ldots\,n}^{a'b'c'\ldots\,n'}
=  S_{AB}(\theta_A - \theta_B)_{ab}^{\alpha b'}
\, S_{AC}(\theta_A - \theta_C)_{\alpha\, c}^{\alpha' c'}
\,\ldots
   S_{AN}(\theta_A - \theta_N)_{\alpha^{(n-2)}n}^{a'\,n'}
\;,
\\[3mm]
  T_B(\theta_A,\ldots,\theta_N)_{a\,\,b\,\,c\,\,\ldots\,n}^{a'b'c'\ldots\,n'}
=  S_{BC}(\theta_B - \theta_C)_{bc}^{\beta c'}
\, S_{BD}(\theta_B - \theta_D)_{\beta\, d}^{\beta' d'}
\,\ldots
   S_{BA}(\theta_B - \theta_A)_{\beta^{(n-2)}a}^{b'\,a'}
\;, \;\; \hbox{etc}
\end{array}
\label{eq:bae3}
\ee
Just as in the two-particle case, we need the eigenvalues $t_A\ldots$
to be pure phases for real rapidities if the multi-particle states are
to have real energy.

If (\ref{eq:usab0}) holds, i.e. if $S_{AB}(\theta)$ is a unitary matrix
for real $\theta$, then it is easy to see that all the \tms\ $T_A$ are
also unitary and hence all their eigenvalues are pure phases.
However, (\ref{eq:usab0}) is not a necessary condition -- if $S_{AB}$
is conjugate to a unitary matrix by a change of basis of the one
particle states, then the \tms\ $T_A$ are also conjugate
to unitary matrices and hence will have phase eigenvalues.
This distinction that the change of basis must only depend on the one
particle states is important: if $S_{AB}$ has phase eigenvalues, then
it is given by a unitary matrix in some basis, but it is only if this
basis is related to the original basis by a change of the one-particle
states that all the higher \tms\ will have phase eigenvalues.

The point of this discussion is that it is very hard to determine
whether a given \sm\ is conjugate to a unitary matrix by a change of
the inner products of one-particle states, since this change of basis
may be very complicated and involve rapidity-dependent changes of
inner products, but if the eigenvalues of any \tm\ are not
phases then we can be sure that this is not the case and that the
finite size spectrum will not be real.

\subsection{Quantisation of affine Toda theory: Quantum group method}
\label{qgquantisation}

As shown by Bernard and LeClair \cite{BLec2}, the `imaginary-coupling'
quantum affine Toda field theory based on the affine algebra $g$ has a
$U_q(g^\vee)$ symmetry. Since this algebra includes the topological
charge as one of its generators, we must assume that the solitons
transform non-trivially under this algebra, and this leads to the
requirement that the soliton-soliton \sms\ are proportional to the
\Rms\ of the symmetry algebra.

The \sm\ of the fundamental excitation is
derived from the universal \Rm\ of an affine quantum group in its
fundamental representation.
There are again two different notations $R$ and $\cR$ for the \Rm\
which are related by $\cR = \bP R$ where again $\bP$ is the
permutation operator.

Such a matrix $\cR(x,q)$ has two
parameters - $q$ the deformation parameter which is a pure phase, and
$x$ the spectral parameter. $\cR$ satisfies the matrix equation
\be
  \cR(x,q) \cR(1/x,q) \propto 1
\;.
\label{eq:ur}
\ee
In the theory of \Rms\ an \Rm\ is also said to be
unitary if (\ref{eq:ur}) holds \cite{smir}.

The \sm\ $\cS_{AA}$ of the fundamental excitation is then obtained
as
\be
  \cS_{AA}(\theta )_{ij}^{kl}
= f(\theta)\,\cR\big(\eta_{AA}\, x(\theta)\,,\,q\,\big)_{ij}^{kl}
\;,
\label{eq:fr}
\ee
where $x(\theta) = \exp( \alpha  \theta )$ for some constant $\alpha$,
$\eta_{AA} = \pm 1$ (the sign to be decided on the basis of the
bootstrap) and where $f(\theta )$ is a suitable scalar function
ensuring (\gup) holds:
\be
  \cS_{AA}(\theta)\, \cS_{AA}(-\theta) = 1
\;,
\label{eq:us11}
\ee
and which can be fixed by requiring crossing symmetry and a
suitable number of CDD poles. (n.b. different physical models may
differ exactly in the number and position of these CDD poles -- see
\cite{Koub4}).

The \sms\ for the higher particles are to obtained using bootstrap
fusion. It is important to note here that since $\cS_{AB}$ and $\cS_{BA}$
are derived {\em independently} using the \sm\ bootstrap, it is not
free for us to impose Parity invariance on a \sm, rather we have to
check whether or not it holds.
As noted by Delius, \gup\ will also hold for all higher
particle \sms, assuming that one can associate a representation
of the quantum group to each particle species \cite{delius}.

As stated before, \gup\ is only equivalent to (\ref{eq:gu}) if the
\sms\ are (P) invariant, and this is usually not the case in
affine Toda field theories -- the only known exception being
Sine--Gordon theory.
If we really want the \sm\ to be a unitary matrix, then
as noted by Smirnov \cite{smir}, the only known method is quantum group
restriction to an RSOS scattering theory of kinks (and breathers).

Quantum group restriction relies on a choice of gradation, and so we
discuss this first.

\subsubsection{Gradations}

The symmetry algebra of $g$--affine Toda field theory is $U_q(g^\vee)$
which is generated by non-local charges with non-trivial Lorentz spin,
and the Lorentz spins of these charges may be changed by
redefining the energy momentum tensor, or, equivalently, by the
addition to the action of a Feigin-Fuchs term as we discuss in the
appendix \ref{app:regrade}.

Furthermore the charges act on multi-particle states by a
representation which is rapidity dependent, the dependence being
determined by the spins of the charges. This may be made clearer by
considering the charges $e_\alpha$ associated to the simple roots
$\alpha$ of $g^\vee$, which have spins
\be
  s_\alpha
= \frac{4 \pi}{\beta^2} |\alpha|^2 - 1 + \gamma\cdd\alpha
\;,
\label{eq:regrade}
\ee
where $\gamma$ is an arbitrary vector corresponding to the ambiguity
in the spins; a choice of $\gamma$ is called a choice of gradation.
There are two distinguished classes of gradations:

\begin{itemize}

\item{} The spin gradation

In this gradation the charges have their canonical spins, and is
given by $\gamma = 0$.

\item{} Homogeneous gradations

A gradation is called homogeneous if all spins $s_\alpha$ are zero
apart from one.

\end{itemize}

The \sms\ depend on the gradation, and a change of gradation is a
rapidity dependent similarity transformation on $S$.

For $U_q(a_2^{(1)})$ all homogeneous gradations are equivalent,
and the algebra of zero-spin charges is $U_q(a_2)$;
there are two inequivalent homogeneous gradations for
$U_q(a_2^{(2)})$, for which the algebras of zero-spin charges are
$U_q(a_1)$ and $U_{q^4}(a_1)$ and which we call the $(1,2)$ and
$(1,5)$ gradations respectively.

\subsubsection{Quantum group restriction}

Quantum group restriction relies on a choice of gradation for which
a subset of the non-local conserved quantities are Lorentz scalars;
this means that the action of the corresponding quantum subalgebra on
multi-particle states commutes with $\cR$.
Given this situation it then makes sense to restrict to highest
weight states of this subalgebra.
As a result, $N$--particle states may be labelled by sequences of
$N+1$ highest weights (obeying certain relations). For $q$ a root of
unity, one can then consistently restrict the Hilbert space further so
that the highest weights of a sequence are chosen from a finite
subset (depending on $q$), and these theories are exactly the quantum
group restricted RSOS theories.
In this formulation we label \sm\ for particles of species $A$ and $B$
in terms of the vacua by
\be
  \epsfbox[0 25 50 50]{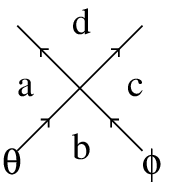}
\;\;\leftrightarrow\;\;
  \kS A,B,a,b,c,d.(\theta-\phi)
\;.
\ee

This procedure preserves several properties of the unrestricted
theory, including the Yang-Baxter relation, \Rm\ unitarity and
crossing symmetry, and furthermore the eigenvalues of the matrices
$t^{(ac)}_{AB}$ whose elements are
\be
  \left(t^{(ac)}_{AB}\right)_{bd} = \kS{A},B,a,b,c,d.
\ee
are a subset of those of $\cS$.
However, what makes this procedure useful and interesting is that the
physically relevant data, i.e. the eigenvalues of the \tms\ (see
\cite{Zamo10,KFV})
\be
  T_{a_1\,a_2\,\ldots a_{k-1}\,a_k}^{b_1\,b_2\,\ldots b_{k-1}\,b_k}
  (\theta_1,\ldots\,\theta_k)
= \delta_{a_1}^{b_2}
  \prod_{j=2}^{k}
  \kS A_1,A_j,b_{j-1},a_j,a_{j+1},b_j.(\theta_1 - \theta_j)
\;,
\label{eq:tm}
\ee
have little relation to those of the unrestricted
\sm\ $S$. This means that it is possible to obtain a unitary theory
from a nonunitary theory by RSOS restriction, and the resulting
theories can have a very rich finite-size spectrum.

%%%%%%%%%%%%%%%%%%%%%%%% section 2 %%%%%%%%%%%%%%%%%%%%%%%%%%%%%%%%

\section{Quantum \sms\ of the $a_2^{(1)}$ theory}
\label{sec:a21quantum}

The first affine Toda theory to be investigated after Sine-Gordon
theory, and the first in which the problems mentioned above may arise,
is the $a_2^{(1)}$ theory.
In the real--coupling case, this leads to a theory of two real scalar
fields which may be thought of as charge conjugates of each other.
The first author to treat the imaginary couplign theory was Hollowood,
who found a spectrum of solitons and breathers, in different species
and different degrees of excitation, with the exact details depending
on the coupling constant $\beta$.

In order to state his results, we give our conventions
(which in this case follow sections 2 and 5 of \cite{smats2})
\be
x = \exp(3 \lambda\theta)
\;,\;\;
q = - \exp(i\pi\lambda)
\;,\;\;
\lambda = \frac{4\pi}{\beta^2} - 1
\;.
\ee
Hollowood found that
the solitons come in two conjugate species $(a) = 1,2$ each with three
flavours (corresponding to the ${\bf 3}$ and ${\bf \bar 3}$
representations of $a_2$), and in various excitation levels
$k = 0,1,2,\ldots,[\lambda]$; such solitons are denoted  $A^{(a)}_k$
and have mass
\be
  M^{(A)}_k
= 2 M \cos\left(\, \frac\pi 3 \left( 1 - \frac{k}\lambda \right)\,\right)
\;.
\ee
The fundamental solitons are then $A^{(1)}_0$ and  $A^{(2)}_0 =
\overline{A^{(1)}_0}$. Hollowood also found that there were two
conjugate breather species $(a)=1,2$  which have excitation levels
$p=1,2,\ldots [ \frac{3 \lambda}2 ]$, are denoted by $B^{(a)}_p$ and have
mass
\be
  M^{(B)}_p
= 2 M \sin\left(\,\frac{\pi p}{3\lambda} \,\right)
\;.
\ee
Again we have $B^{(2)}_p = \overline{B^{(1)}_p}$.

Hollowood only considered the \sms\ for the scattering of fundamental
solitons and anti-solitons, and the first author to treat the
remaining \sms\ was Gandenberger \cite{Georg1}. We give their results
and comment upon them in the next two sections.

\subsection{%
The unrestricted \sms\ for soliton--soliton scattering in \ato}
\label{sec:a21ss}

The unrestricted \sms\ for \ato\ soliton scattering are given in
\cite{Holl1}, but the eigenvalues were not investigated, nor was the
matrix-unitarity of the \sm.
The relevant \sms\ are the $9{\times}9$ matrices $S_{A_k\,A_l}(\theta)$ and
$S_{A_k\,\barA_l}(\theta)$, which take the form
\be
  S_{A_k\,A_l}(\theta) =
  f_{kl}(\theta)\,R_{{\bf 33}}(\eta_{kl}\,x,q)
\;,\;\;\;\;
  S_{A_k\,\barA_l}(\theta) =
  \tilde f_{kl}(\theta)\,R_{{\bf 3\bf \bar 3}}(\tilde\eta_{kl}\,x,q)
\;,\;\;\;\;
\ee
where the prefactors $f_{kl}, \tilde f_{kl}$ are
always phases for real $\theta$ and
$\eta_{kl}$ and $\tilde\eta_{kl}$ are phases which may depend on the
gradation. In the homogeneous and spin gradations these phases take
values $\eta_{kl} = -\tilde\eta_{kl} = (-1)^{k+l}$.

The eigenvalues of $R_{\bf 33}(x,q)$ are each triply degenerate, and are
\be
1
\;,\;\;\;\;
\left(\frac{ 1 + q \sqrt x  }{q + \sqrt x  } \right)
\;,\;\;\;\;
\left(\frac{ 1 - q \sqrt x  }{q - \sqrt x  } \right)
\ee
which are clearly phases for non-negative real $x$ and $q$ a phase.
However, for $q$ a phase and $x$ real and negative these are
not phases, and so we immediately see that the \sm\ $S_{A_0A_1}$ has
non-physical eigenvalues. Hence for the unrestricted theory to be
unitary, we need to be in the regime $\lambda<1$ for which there are
no excited solitons (Hollowood also reached this conclusions but from
different reasonings).

Furthermore, when we investigated the Bethe-Ansatz equations for three
particles of type $\bf 3$ numerically, we found that the \tms\ did not
in general have phase eigenvalues. Hence, while $S_{\bf 33}$ is
conjugate to a unitary matrix, this does not simply involve a change
of inner product on the one particle states, and even the fundamental
$S_{A_0 A_0}$ \sm\  is not well behaved.

The problems with \ato\ are also clearly seen in the eigenvalues of
$R_{{\bf 3\bar 3}}(x,q)$, which are 1 (six times degenerate) and
the three eigenvalues
\be
\left(\frac{ 1 - q \,x^{1/3} }{q - x^{1/3}} \right)
\;,\;\;
\left(
  \frac{1-q\,x^{1/3}\,e^{2\pi i/3}}{q-x^{1/3}\,e^{2\pi i/3}}\right)
\;,\;\;
\left(
  \frac{1-q\,x^{1/3}\,e^{4\pi i/3}}{q-x^{1/3}\,e^{4\pi i/3}}\right)
\;.\;\;
\ee
Thus the eigenvalues of $R_{\bf 3\bar 3}(x,q)$ are not all phases for
$q$ a generic phase and $x$ a positive {\em or} negative real number,
and there is no need to examine the three-particle states.

This certainly suggests (contrary to the comments in \cite{Holl1})
that the unrestricted \ato\ theory is not a unitary theory (in any
sense other than \gup), even in
the repulsive (no bound state) regime, apart from possible discrete
values of the coupling constant.

\subsection{%
The \sms\ for soliton--breather scattering in \ato}
\label{sec:a21sb}

While Hollowood discussed the presence of breathers in the spectrum,
Gandenberger was the first to compute their \sms;
his results for the \sms\ of the fundamental soliton ($A {=} A^{(1)}_0$),
anti--soliton ($\barA$), fundamental breather ($B {=} B^{(1)}_1$) and
conjugate breather ($\barB$) are
\bea
&& S_{\vphantom{\barA}AB} = S_{\barB A} = S_{\barA\, \barB} = S_{B\barA} =
\;,
\frac{ \lb{ 3 + B} \lb{ 3 - B} }
     { \lb{-1 + B} \lb{ 7 - B} }
\nonumber\\
&& S_{\vphantom{\barA}BA} = S_{\barA B} = S_{\barB\,\barA} =S_{A \barB} =
\frac{ \lb{ -7 + B} \lb{ 1 - B} }
     { \lb{ -3 - B} \lb{-3 + B} }
\;,
\label{sa21}\eea
where
\be
 \lb x
= \sinh\left( \frac{\theta}{2} + \frac{i\pi\,x}{12} \right)
\;,\;\;\;\;
 B = -\frac{1}{2\pi} \frac{ \beta^2}{1 - \beta^2/4\pi}
   = - \frac 2\lambda
\;.
\ee
As can be readily seen, this does not satisfy the full range of
properties, breaking (P) and (T), but preserving $(A)$, $(C)$ and
$(PT)$ while also being crossing symmetric.
Also, since it is derived by the quantum--group method, it obeys
\gup\ (\ref{eq:gu2})
\be
  S_{AB}(\theta)\, S_{BA}(-\theta) = 1
\;,
\ee
for all $\theta$, but since it breaks $(P)$, it does not satisfy
(\ref{eq:gu}), that is for real $\theta$,
\be
  S_{AB}(\theta)\, \left({S_{AB}(-\theta)}\right) \neq 1
\;,
\ee
Hence, in this case, we can attribute the failure of unitarity
(\ref{eq:usab0}) to the failure of parity, since unitarity is
equivalent to (A), \gup\ and (P), and of these only the last fails.
As a result, we expect that this model will have a complex spectrum
whenever it contains both breathers and solitons.

%%%%%%%%%%%%%%%%%%%%%%%% section 3 %%%%%%%%%%%%%%%%%%%%%%%%%%%%%%%%

\section{Classical scattering in $a_2^{(1)}$}
\label{a21classical}

In the previous section we have just seen that the quantum scattering
of solitons with breathers in $a_2^{(1)}$ has severe problems -- but
how can this be reconciled with the supposedly consistent scattering
in the classical theory?
To answer this question we first have to recall some elements of the
construction of soliton and breather solutions.

All the known soliton solutions of the $a_n^{(1)}$ Toda
theories were found by Hollowood in \cite{Holl3}, the breathers by
Harder et al in \cite{HIMc1}, and the scattering of solitons by Olive
et al. in \cite{FJKO1}. We take the action of \ato\ to be
\be
  S
 = \frac 12( \dot\phi^2 - \phi'^2 )
\;-\; \frac{\mu^2}{\beta^2} \sum_{i=0}^2
   \left( \exp( {\rm i}\,\beta\alpha_i \cdot \phi ) - 1 \right)
\;,
\label{eq:act1}
\ee
where $\beta$ is imaginary, and $\alpha_i$ are the simple roots of
\ato, with $\alpha_i^2 = 2$.
The stationary points of the potential in (\ref{eq:act1}) are
\be
  \phi
= \frac{ 2 \pi }\beta\,( \,m_1 \lambda_1 + m_2\lambda_2\,)
\;,
\label{eq:vacua}
\ee
where $\lambda_i$ are the fundamental weights of \ato. We consider
only the solutions which tend to one of these vacua as $|x|\to\infty$,
and denote the topological charge of such a solution by
$(m_1,m_2)$, where
\be
  \phi|_{x \to \infty}- \phi|_{x \to -\infty}
= \frac{ 2 \pi }\beta\,( \,m_1 \lambda_1 + m_2\lambda_2\,)
\;.
\label{eq:tch}
\ee
Hollowood gives the generic multi-soliton solution of $a_2^{(1)}$ in
terms of tau-functions $\tau_j$ as
\be
\begin{array}{rcl}
  - \beta \phi
&\!\!\!=\!\!\!&\ds
 \sum_0^2 \alpha_j \log \tau_j
= \frac{1}{\sqrt 2}\left( \log(\tau_1^2/(\tau_0\tau_2)),
                          \log(\tau_2/\tau_0) \right)
\;,\\[2mm]
  \tau_j
&\!\!\!=\!\!\!&
\ds
\sum_{k=0}^N\;
\sum_{ 1 \leq j_1 < \ldots < j_k \leq N}
  \left( \prod_{i=1}^k
         \omega^{  j a_{j_i} } p_{j_i} X(\theta_{j_i},x,t) \right)
  \left( \prod_{1 \leq i < i' \leq k }
         X_{a_{j_i}, a_{j_{i'}}}( \theta_{j_i} - \theta_{j_{i'}} )
  \right)
\;,
\end{array}
\label{eq:hol}
\label{eq:tj1}
\ee
where
\be
{\renewcommand{\arraystretch}{1.4}
\begin{array}{c}
  \om
= \exp( 2 \pi i / 3 ) \;,\;\;
   X(\theta,x,t)
= \exp( m( x \cosh \theta - t \sinh\theta)) \;,\\[2mm]
  X_{1,1}(\theta) = X_{2,2}(\theta)
= {\ds \frac{ \cosh\theta - 1 }{\cosh\theta + 1/2}} \;,
\;\;\;\;
  X_{1,2}(\theta) = X_{2,1}(\theta)
= {\ds \frac{ \cosh\theta - 1/2 }{\cosh\theta + 1}} \;.
\end{array}
}
\ee
where $m = \sqrt 3\,\mu$, and a particular solution is given by a
choice of $N$ and the $3 N$ parameters $\{a_i,\theta_i,p_i\}$ with
$a_i \in \{1,2\}$,  $\theta_i\in \mathbb C$ and $p_i \in \mathbb C^*$; such
a solution has total energy $E$ and momentum $P$ given by
\be
  E = M \sum_i \cosh \theta_i
\;,\;\;\;\;
  P = M \sum_i \sinh \theta_i
\;,
\label{eq:EP}
\ee
where $M = (8 m)/\beta^2$.
To find the allowed values of $p_i$, $\theta_i$, $a_i$ for which this
gives a solution regular for all $x$, $t$, is in general a difficult problem,
as it is to find the overall topological charge of a solution, but
this can be  fully analysed for the cases $N{=}1$ and $N{=}2$.

\subsection{The fundamental soliton solutions of \ato}

The simplest such solutions are the single solitons, which
are parametrised by a rapidity $\theta$, a `species' label
$a\in\{1,2\}$, and a complex
number $p$ of which the magnitude determines the initial position and
the phase determines the `shape' and (together with $a$) the
topological charge. For the single solitons, (\ref{eq:tj1}) becomes
\be
  \tau_j =
1 + \omega^{aj} p X(\theta,x,t)
\;.
\ee
This solution will be singular for any values of $x,t$ for which one
of $\tau_j$ is zero. Since $X(\theta,x,t)$ can take any real positive
values,  each $\tau_j$ will be zero  for some (particular) values of
$x$ and $t$ if the phase of $p$ is $\pi + 2n\pi/3$ for any integer $n$.
As a result, the only restriction is on $p$ which is that its phase
must not be equal to $(2n + 1)\pi/3$ for any integer $n$.

For each choice of $a = 1,2$, the topological charge of the soliton
is constant on each region of allowed $p$ values.
By inspection, we find that the topological charges (\ref{eq:tch})
take the values $(m_1,m_2)$ as shown in figure \ref{fig:tch1}.
Note that $a=1$ gives topological charges in the fundamental
representation $(1,0) \equiv {\bf 3} $ of $a_2$, and $a=2$ in the
fundamental representation $(0,1) \equiv {\bf \bar 3}$.

{
\begin{figure}[ht]
\[
\begin{array}{cc}
\epsfxsize=3cm
\epsfbox[0 0 288 288]{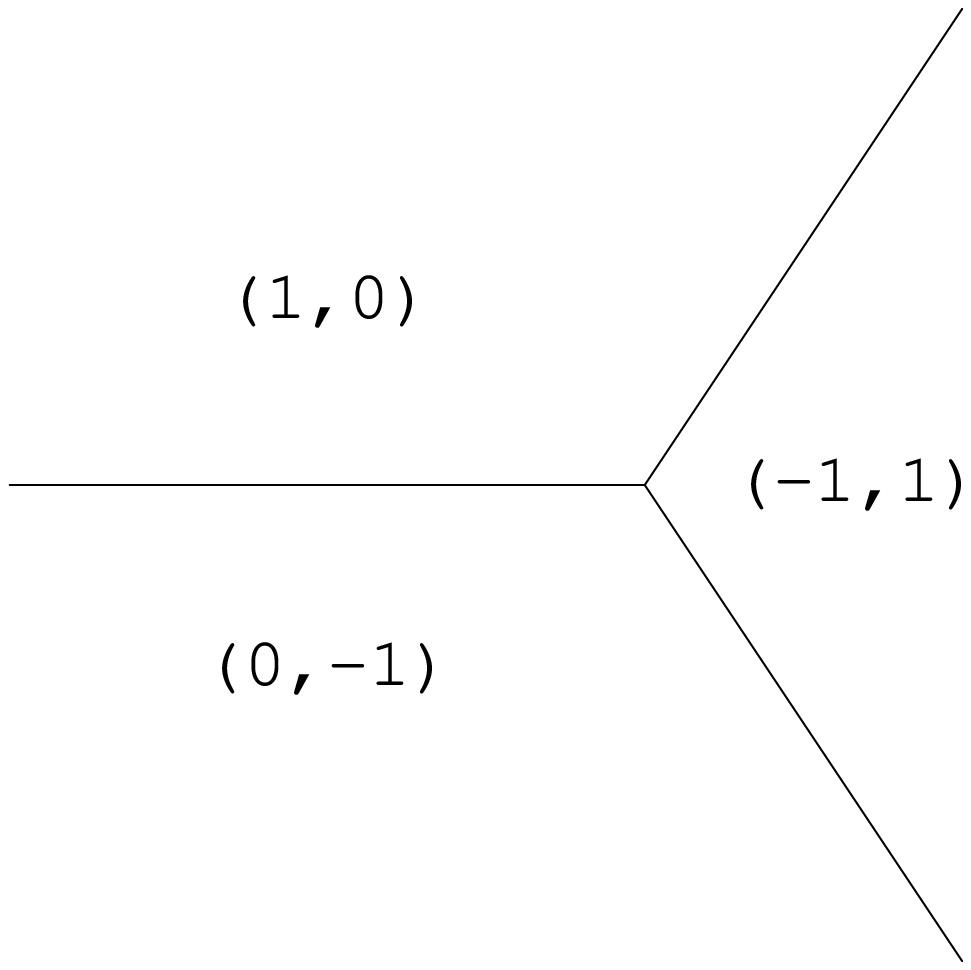}
{}~~~~&~~~~
\epsfxsize=3cm
\epsfbox[0 0 288 288]{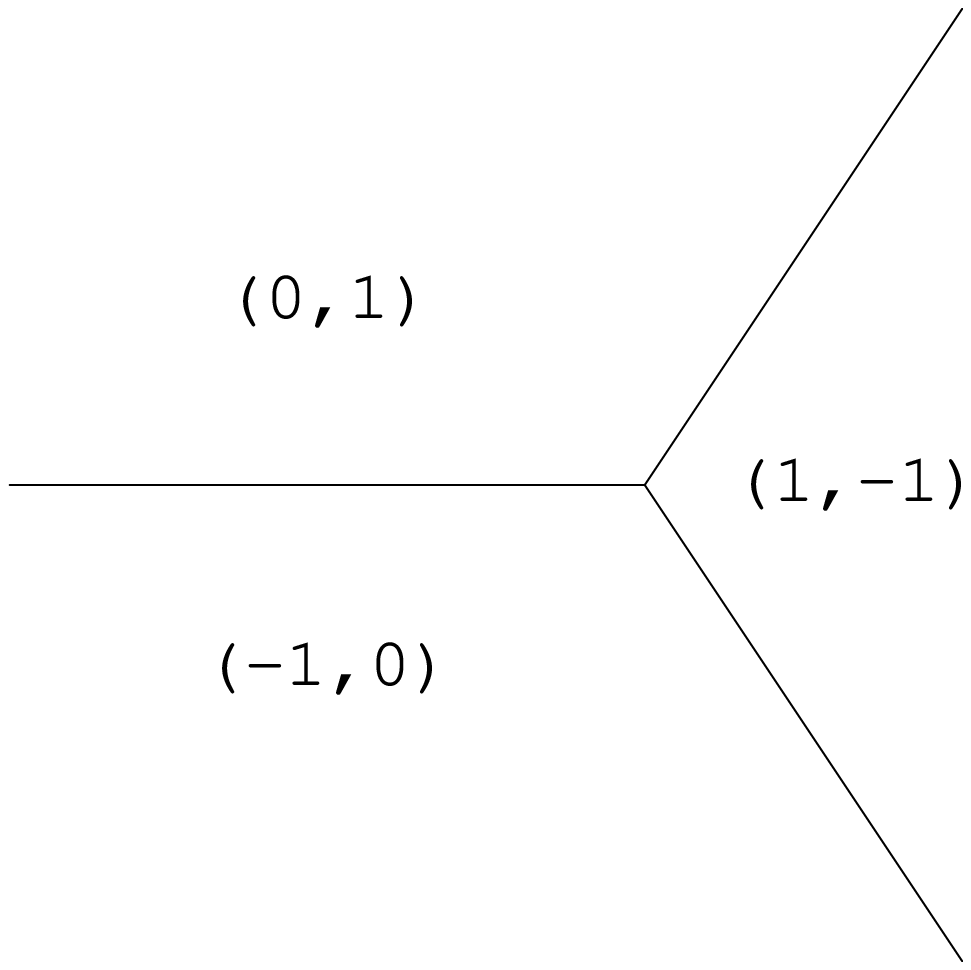}
\\
  \hbox{$a = 1$}
&
  \hbox{$a=2$}
\end{array}
\]
\caption{\small
The dependence on $p$ of the topological charges of the fundamental
solitons of \ato\ for $a=1$ and $a=2$.}
\label{fig:tch1}
\end{figure}
}

\subsection{Solutions with $N=2$}

For a solution (\ref{eq:hol})
of \ato with $N=2$ to have real energy and momentum,
we must either take $\theta_1, \theta_2$ to be both real,
or to be a complex conjugate pair.
The solutions with two real rapidities correspond to scatterings of two
fundamental solitons and we discuss these in section \ref{sec:scatter}, and
now we consider only the solutions with complex conjugate rapidities.
We also can simplify further our analysis by a Lorentz transformation
so that $\theta_1 {=} {-} \theta_2 {=} i\,\alpha$ is pure imaginary and the
solutions are now stationary and periodic in time.
The mass of such a solution is $2 M \cos \alpha$, where $M$ is the
mass of a single soliton, and consequently the fundamental region of
$\alpha$ is $[-\pi/2,\pi/2]$.

By the following redefinitions,
\be
\begin{array}{rclrcl}
 p_1 &\!\!=\!\!& r_1 \exp( i \phi_1 )\;,\;\;\;\;
&
 p_2 &\!\!=\!\!& r_2 \exp( i \phi_2 )\;,
\\[2mm]
 u        &\!\!=\!\!& \sqrt{ r_1 / r_2 }\;,
&
 \delta_j &\!\!=\!\!&  (\phi_1 + \phi_2) / 2
               \,+\,
               \pi j ( a + b ) / 3\;,
\end{array}
\label{eq:zj}
\ee
\be
  z_j(x,t)
{}~=~ \sqrt{ r_1 r_2}
  \, X(i \alpha,x,t)
  \, \exp( \pi i j (a - b) / 3 )\;,
\ee
the general two soliton solution with parameters $\{a,p_1,i \alpha\}$ and
$\{b,p_2,-i\alpha\}$ can be cast in the form
\be
  \tau_j
= 1
\,+\,
  e^{ i \delta_j} \left( u\,z_j(x,t)  + z_j(x,t)^* / u \right)
\,+\,
  e^{ 2 i \delta_j } X_{a,b}(2 i \alpha)\,| z_j(x,t) |^2
\;.
\label{eq:tj2}
\ee
The $x$ and $t$ dependence of (\ref{eq:tj2}) is contained in
$z_j(x,t)$, which will take any value in $C^*$ for suitable $x$ and $t$.
Hence, the two-soliton solution will be singular if
\be
  \tau_j
= 1
\,+\,
  e^{ i \delta_j} \left( u\,z  + z^* / u \right)
\,+\,
  e^{ 2 i \delta_j } X_{a,b}(2 i \alpha)\,| z |^2
\;,
\label{eq:tj3}
\ee
is zero for any complex number $z$. This is amenable to complete
analysis, which we relegate to appendix \ref{app:sings}, and the result
is that the solution is singular for some value of $x$ and
$t$ provided that, for each $j$,
\bea
  \Big(
  \tau  - ( 2 X_{ab}(2 i \alpha)(\cos(2 \delta_j)+1) - 2 )
  \Big)
  \Big(
  \tau  - ( 2 X_{ab}(2 i \alpha)(\cos(2 \delta_j)-1) + 2 )
  \Big)
& \geq & 0\;, \\
  2 \,\Big( 2 X_{ab}(2 i \alpha) - 1 \Big)\, \cos(2 \delta_j)
& \leq &
  \tau \;,
\eea
where $\tau = u^2 + u^{-2}$.

To find zero-topological charge breathers we must
take $a=1$ and $b=2$, and find
$X_{12}(2 i \alpha) = 1 - 3 / (4 \cos^2 \alpha)< 1$.
Since we require $X < 0$ or $X>1$, we must take $\pi/2>|\alpha| > \pi/6$, so
that $0<M(\alpha) < \sqrt 3 M$. We note here that the solutions with
$\alpha>0$ and $\alpha<0$ can be regarded as conjugate to each other,
in the manner of Gandenberger.
It is also interesting to note that the maximum mass of a regular
classical breather solution is $\sqrt 3\, M$, whereas quantum
considerations suggest the breathers have masses up to $2 \,M$.
Whether this signals that the quantum \sms\ obtained from quantum
groups methods are not the quantisations of the classical theories,
or that there are classical breather solutions to be found, or whether
there is some other explanation, we cannot say.

Next, since $\delta_j(x,t)$ is independent of $j$,
we need only consider $j=0$ with the result that for each value of
$\pi/6<\alpha<\pi/2$ there  is a single finite region of
$\{2<\tau,0<\delta<\pi\}$ with regular solutions, bounded by the
curves
$\tau = 2$ and
$\tau = 2 + 2 X_{12}(2i\alpha)(\cos(2\delta)-1)$.
For example,
figure \ref{fig:b1} shows the allowed region for $\alpha = \pi/3$.
By inspection, the topological charge of a regular breather solution
is always zero.
{
\begin{figure}[ht]
\[
\epsfxsize=7cm
\epsfbox[0 50 288 238]{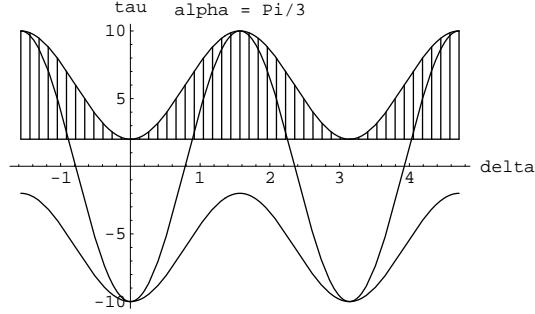}
\]
\caption{
Allowed ranges of $\tau,\delta$ (shaded) for a breather of
$a_2^{(1)}$ with $\alpha = \pi/3$
}
\label{fig:b1}
\end{figure}
}

\subsection{Scattering of two solitons in $a_2^{(1)}$}
\label{sec:scatter}

The scattering of solitons has already been investigated in
\cite{FJKO1}. The scattering process itself is complicated, but it is
quite straightforward to examine the final state,
and it was shown that the only result of solitons scattering is a change
in trajectory of the participating solitons.

If we want to describe the scattering of a fundamental soliton of
parameters $\{a_1,\theta_1,p_1\}$ with another fundamental soliton of
parameters $\{a_2,\theta_2,p_2\}$, this is not simply the solution
(\ref{eq:hol}) with parameters $\{a_1,\theta_1,p_1;a_2,\theta_2,p_2\}$.
To find the correct $N=2$ soliton solution we must consider the $N=2$
solution with parameters $\{a_1,\theta_1,q_1;a_2,\theta_2,q_2\}$
so that $\tau_j$ takes the form
\be
 \tau_j
 = 1 + \om^{a_1 j} q_1 Y + \om^{a_2 j} q_2 \Upsilon +
    \om^{(a_1+a_2)j} X_{a_1,a_2}(\theta_{12}) q_1 q_2 Y \Upsilon
\;,
\ee
where
\be
 Y = X(\theta_1,x,t)
\;,\;\;
 \Upsilon = X(\theta_2,x,t)
\ee
We assume that $\theta_{12}$ is positive,
so that in the initial state for $t \ll 0$, $|Y| \gg  |\Upsilon|$.
As a result, for the two ranges of $x$ such that $|Y q_1 |\sim 1$ and
$|\Upsilon q_2|\sim 1$, the solution describes two well
separated solitons. In the first case, $|q_2 \Upsilon|\ll 1$,
so that for this range of $x$ the tau function is dominated by
\be
\tau_j
\sim
 1 + \om^{a_1 j} q_1 Y
\;,
\ee
which is a single soliton of parameters $(a_1,\theta_1,q_1)$.
However, in the range of $x$ such that $|\Upsilon q_2|\sim 1$, we see
that $|Y q_1 |\gg 1$ so that the tau function is dominated by
\be
\tau_j
\sim
 q_1 Y \om^{a_1 j}
 \left( 1 + \om^{a_2 j} (q_2 X_{a_1,a_2}(\theta_{12})) \Upsilon \right)
\;,
\ee
which is a single soliton with parameters
$(a_2,\theta_2,q_2 X_{a_1,a_2}(\theta_{12}))$.
Hence, in the initial configuration, we must take the parameters $q_1$
and $q_2$ to be
\be
  q_1 = p_1
\;,\;\;\;\;
  q_2 = \frac{p_1}{X_{a_1\,a_2}(\theta_{12})}
\;.
\ee

We can repeat this exercise for $t \gg 0$ and find that the solitons
in the final state have parameters
\be
 ( a_1,\theta_1,q_1 X_{a_1,a_2}(\theta_{12}))
\;,\;\;\;\;
 ( a_2,\theta_2, q_2)
\ee
Thus it appears that the net effect of the scattering process has been
to change the parameters of the two solitons in the following way:
\bea
 (a_1,\theta_1, p_1 ) &\mapsto&  ( a_1,\theta_1,p_1 X_{a_1,a_2}(\theta_{12}))
\nn
 (a_2,\theta_2, p_2 ) &\mapsto&  ( a_2,\theta_2,p_2 /X_{a_1,a_2}(\theta_{12}))
\eea
However, since $X_{a_1,a_2}(\theta_{12})$ is real for real rapidities
$\theta_i$, the net effect can be simply absorbed into a time delay or
advance for the trajectory of each soliton, as can be seen from
\be
  \tau_j(a_1,\theta_1,p_1 X_{a_1,a_2}(\theta_{12}) )(x,t)
= \tau_j(a_1,\theta_1,p_1 )(x,t + \Delta t_1)
\;,
\ee\be
  \tau_j(a_2,\theta_2,p_2/X_{a_1,a_2}(\theta_{12}) )(x,t)
= \tau_j(a_2,\theta_2,p_2 )(x,t + \Delta t_2)
\;,
\ee
where
\be
 \exp( m \sinh \theta_1\,\Delta t_1 ) = X_{a_1,a_2}(\theta_{12})
\;,\;\;\;\;
 \exp( m \sinh \theta_2\, \Delta t_2 ) = X_{a_1,a_2}(\theta_{12})^{-1}
\;.
\ee
There is perhaps some slight sign of potential problems in the quantum
theory in that for each choice of $\theta_1,\theta_2,p_1$,
then for any $x,t$ one can find a value of $p_2$ such that the solution is
singular at that point -- but in general these singularities are
isolated in spacetime, and one might argue that they present no real
problems.

\subsection{The results of general scatterings in \ato}

It should be clear from the previous calculation how to generalise
this result to the arbitrary scatterings of solitons, breathers and
breathing solitons of \ato, simply by considering the scattering of their
constituent solitons.

The situation of interest to us is the scattering of a breather with
a single soliton. Consider a soliton of parameters $(a,\theta,p)$
scattering from a breather of parameters $(\phi,\alpha,q_1,q_2)$,
which in terms of its constituent solitons is formed from two solitons of
parameters
$(1,\phi+i\alpha,q_1)$ and $(2,\phi-i\alpha,q_2)$, with $\theta>\phi$
as before.

After scattering, the labels and rapidities are unchanged but the
`shape' parameters have changed as follows:
\bea
  p &  \mapsto &
  p X_{a1}(\theta - \phi - i\alpha) X_{a2}(\theta - \phi + i \alpha )
\\
 q_1 & \mapsto &
 q_1 / X_{a1}(\theta - \phi - i\alpha)
\\
 q_2 & \mapsto &
 q_2 /  X_{a2}(\theta - \phi + i \alpha )
\eea

If we confine our attention to the breather, we see that the result of
the scattering is not simply a change in trajectory.
The `shape' parameter of the breather has changed by
\be
  u^2 \exp(2 i \delta) = q_1 / q_2^*
\mapsto
  u'^2 \exp(2 i \delta') =
  u^2 \exp(2 i \delta)
\frac{ X_{a2}(\theta - \phi + i \alpha )^* }
     { X_{a1}(\theta - \phi - i\alpha) }
\ee
For $a=1,2$ this factor takes the values
\be
\frac{ X_{a2}(\theta - \phi + i \alpha )^* }
     { X_{a1}(\theta - \phi - i\alpha) }
 = \left\{
\begin{array}{rr}
\ds
\frac{ (\cosh( \theta - \phi - i \alpha) - 1/2)
       (\cosh( \theta - \phi - i \alpha) + 1/2)  }
     { (\cosh( \theta - \phi - i \alpha) + 1 )
       (\cosh( \theta - \phi - i \alpha) - 1 )   }\;,  & a=1 \\[5mm]
\ds
\frac{ (\cosh( \theta - \phi - i \alpha) - 1)
       (\cosh( \theta - \phi - i \alpha) + 1)    }
     { (\cosh( \theta - \phi - i \alpha) + 1/2 )
       (\cosh( \theta - \phi - i \alpha) - 1/2 ) }\;,  & a=2
\end{array}
\right.
\ee
For generic values of $\theta,\phi$ and $\alpha$ this is a generic
complex number, so that in a typical scattering both $\delta$ and $u$
will be altered. Note however that this factor tends to one as
$|\theta-\phi|\to\infty$.

We can now try as before to absorb all the effects of the scattering
into an effective change of trajectory, corresponding to a
time displacement $\delta t$ and space displacement
$\delta x$. From eqn.\ (\ref{eq:tj2}) this leads to the two equations
\bea
  (q_1  / X_{a1}(\theta - \phi - i\alpha) )
\! &=& \!
  q_1
  \exp(m (\delta x \cosh(\phi+i\alpha) - \delta t \sinh(\phi+i\alpha)))
\nn
  (q_2  / X_{a2}(\theta - \phi + i\alpha) )
\! &=& \!
  q_2
  \exp(m (\delta x \cosh(\phi-i\alpha) - \delta t \sinh(\phi-i\alpha)))
\label{eq:z1}
\eea
The result is that both $\delta x$ and $\delta t$ are {\em complex}
for a generic scattering, as is the effective time delay
\be
 \Delta T = \frac{\delta x }{v} - \delta t
\;.
\label{eq:z2}
\ee
While this may appear a strange procedure from the classical point of
view, it agrees perfectly with the results of Gandenberger, as the
time delay $\Delta T$ for a light breather scattering off a heavy
soliton this can be compared with the quantum scattering using the
formalism of Faddeev and Korepin \cite{FKor1}, and we find perfect
agreement, as we now show.

Let us recall how the correspondence between the phase-shift and the
classical time delay is made. We consider the scattering of the lowest
breather with mass $m_{B_1}$ in the $a_2^{(1)}$ theory on a soliton of
mass $M$ at rest. The energy in this frame is
$E=M+m_{B_1}\cosh\theta$, where  $\theta $ is the rapidity of the
incoming breather. Given the phase-shift $\delta(E)$ as a function of
the incident energy $E$ the classical time delay can be computed as
\begin{equation}
  \Delta T(E)
= \frac{d\delta(E)}{dE}
\;,
\label{eq:z3}
\end{equation}
assuming that $M \gg m_{B_1}$ which means that we treat the soliton
as a static potential well. In terms of the rapidity we can write
\begin{equation}
  \Delta T(E)
=  \frac{1}{\sinh\theta}\frac{d}{d\theta }
   \left(-i\log S^{\vphantom\phi}_{\!AB}(\theta)\right)
\;,
\end{equation}
where $S_{AB}(\theta)$ is given by eqn. (\ref{sa21}). The
semi-classical limit is  given by taking $\beta\rightarrow
0$. Performing the calculation we find
\begin{equation}
  \Delta T_{limit}(E)
= \frac{\cosh\theta}{m\sinh\theta}\left(\frac{1}{i\sinh\theta -1}-
  \frac{1}{i\sinh\theta +1/2}\right)
\;.
\label{sclimit}
\end{equation}
We only need to check now that in the semi-classical limit the
breather mass is much less than the soliton mass:
the semi-classical limit is $\beta \to 0$, i.e. $\lambda \to \infty$,
so that $m_{B_1} \sim \frac{\pi 2 M}{3\lambda} \sim \frac{\beta^2
M}{6} = m$, and indeed
$m_{B_1} \mathop{\rightarrow}\limits_{\beta\rightarrow 0} m$ and
the approximation $M \gg m_{B_1}$ is valid in the semi-classical limit.

If we now use the results (\ref{eq:z1}) and (\ref{eq:z2}) in
(\ref{eq:z3}),  we can compute $\Delta T$ for a breather
of internal momentum
\begin{equation}
\alpha=\pi/2-\epsilon\ ,\ \epsilon\rightarrow 0\ ,
\end{equation}
which corresponds to the semi-classical limit of the first breather $B_1$.
The result is
\begin{equation}
\Delta T_{classical}(E)=
\frac{\cosh\theta}{m\sinh\theta}\left(\frac{1}{i\sinh\theta -1}-
\frac{1}{i\sinh\theta +1/2}\right)\ ,
\end{equation}
which coincides with (\ref{sclimit}).
If instead we had taken $\alpha = -\pi/2 + \epsilon$, we would have
instead found agreement with $S_{A\barB}$.

Therefore one concludes that the pathological behaviour of the classical
breather-soliton scattering, i.e. the existence of complex time delays
precisely corresponds to the non-unitarity of the phase shift observed
at the quantum level and that the quantum scattering of solitons and
breathers in \ato\ is fundamentally flawed.

Moreover, on closer examination we find that the scattering is also
fundamentally flawed at the classical level. Since the effect of
breather--soliton scattering is to change the shape parameters of both
solutions, there is no guarantee that the final shape parameters will
be in the allowed ranges for the solutions to be regular. While it is
not hard to find explicit values which lead to singular final states
from regular initial states, the following argument is quite general
and illustrates the point well.

Consider the scattering of a fast soliton off a
breather. This will change the parameter $\delta$ of the breather by a
small amount $\Delta\delta$ . Then, $n$ repeated scatterings of the
same species of soliton with very similar rapidities (exactly equal
rapidities are not allowed) off the breather will change $\delta$ by
$n \Delta\delta$, and  will eventually be forced into the singular region.
Thus, the final state obtained starting from an initial state of
regular solitons and a regular breather will be solitons and a
singular breather.

%%%%%%%%%%%%%%%%%%%%%%%% section 4 %%%%%%%%%%%%%%%%%%%%%%%%%%%%%%%%

\section{Quantum \sms\ of the $a_2^{(2)}$ theory}
\label{a22quantum}

The $a_2^{(2)}$ action has a single scalar field, and the unreduced
\sm\ was first given by Smirnov \cite{smir}.
This \sm\ describes the scattering of a soliton transforming in the
triplet representation of $U_q(a_2^{(2)})$, that is three particles of
topological charge $-1, 0$ and $1$ units respectively.

As we mentioned before, $U_q(a_2^{(2)})$ has two inequivalent
subalgebras,  $U_q(a_1)$ and $U_{q^4}(a_1)$, which means that there
are two inequivalent ways of performing the RSOS restriction, which we
refer to as the (1,2) and (1,5) restrictions, the RSOS reduced \sms\
corresponding to perturbations of Virasoro minimal models by these two
primary fields.

\subsection{Unrestricted \sms\ of \att}

In any gradation, the unrestricted \sm\ of the fundamental
soliton $K_0$ is given by
\be
  S_{K_0 K_0}(\theta)_{ij}^{kl}
= S_0(\theta) \,\, R(x,q)_{ij}^{kl}
\label{eq:SmatZMS}
\ee
where $q$ and the spectral parameter $x$ are given in terms of a
variable $\xi$ parametrising the coupling constant dependence of the
theory by%
\footnote{The parameter $q$ appearing in this \Rm\ does not agree with
the conventions of \cite{smats2}. Depending on the definition of the
quantum group relations used, the definition of $q$ depends
on the lengths of the roots of the Lie algebra or on the
numerical values of the symmetrised Cartan matrix. For an explanation
of the conventions used in this section, see \cite{takacs}.}
\be
 x = \exp\left(2\pi\theta/\xi\right)
\;,\;\;
 q = i\exp\left(\frac{i\pi^2}{3\xi}\right)
\;.
\label{eq:xi}
\ee

In the $(1,2)$ gradation, if we define the matrix $\cR = \bP R$
where $\bP$ is the permutation matrix on the two spaces,
then $\cR$  commutes with the
$(1,2)$ subalgebra of $U_q(a_2^{(2)})$ and can be expanded in terms of
projectors $\cP$ on the irreducible representations of the $U_q(a_1)$
algebra in the tensor product
${\bf 3}\otimes{\bf 3} = {\bf 1}\oplus {\bf 3}\oplus {\bf 5}$ as
\be
  {\cR}(x,q)
= {\cP}_{\bf 5}+
  \frac{xq^4-1}{x-q^4}{\cP}_{\bf 3}+
  \frac{xq^6+1}{x+q^6}{\cP}_{\bf 1}
\;.
\label{eq:pexpand}
\ee
While it is obvious that the eigenvalues of $\cR$ are phases for real
$x$ and $q$ a phase, the physically relevant eigenvalues are those of
$R$, which are only phases for $x>0$ and $q$ a phase; for $x<0$ and
$q$ a phase they are not.

To be explicit, the eigenvalues of $R(x,q)$ are three pairs of doubly
degenerate  eigenvalues,
\be
1
\;,\;\;\;\;
\left(\frac{ 1 - q^2 \sqrt x }{q^2 -  \sqrt x } \right)
\;,\;\;\;\;
\left(\frac{ 1 + q^2 \sqrt x }{q^2  + \sqrt x } \right)
\ee
and three eigenvalues $\lambda$ satisfying
\be
{\renewcommand{\arraystretch}{1.4}
\begin{array}{rl}
   (x - q^4)(x + q^6)  \lambda^3
\;+\;  q^6( 2  + q^2 )( \lambda^2 + x^2 \lambda)
&\\
+\; x\,(q^2 - 1)(1 - 3 q^4 + q^8)
   ( \lambda^2 + \lambda)
&\\
-\;  q^2(  1 + 2 q^2) ( x^2 \lambda^2 + \lambda)
\;+\;   (1 - q^4 x)(1 + q^6 x)
&= 0\;.
\end{array}
}
\ee
The \Rm\ in the (1,5) gradation differs by a similarity
transformation from the previous one:
\be
  \tR(y,q)
= y^{H_2}\,R(y^2,q)\,y^{-H_2}
\;,
\label{regr}
\ee
where for convenience we introduced a new spectral parameter
$y = \exp\left(\pi\theta/\xi\right)$, an element $H$ in the Cartan
subalgebra of $U_q(a_2^{(2)})$, and where the subscript $2$ indicates
that this matrix acts in the second space. Explicitly,
\be
  H
= \left(\matrix{ 1&0&0\cr 0&0&0\cr 0&0&-1}\right)
\;.
\label{eq:h2}
\ee
The eigenvalues of $\bP \tR$ are different from those of $\cR$, but
the eigenvalues of the \sms\ are unaffected by the regradation as this
is just a similarity transformation on $R$.

{}From (\ref{eq:pexpand}) it is clear that there are poles in
$S_{K_0K_0}$ for $x=q^4$ and $x=-q^6$. The latter correspond to the
possibility to have scalar
(breather) bound states $B_p$
and the former to excited solitons $K_m$. The exact spectrum may be
very complicated (see \cite{smir}) but at least for these particles we
have
$k = 0,1,2,\ldots,[\lambda]$, of mass%
\footnote{It is clear that these are exactly the same masses as in the
\ato\ theory with $\lambda = 2\pi/(3\xi)$, but there is no other
obvious relation between \sms\ of the two theories.}
\be
  M( K_m )
= 2 M \cos\left(\, \frac\pi 3 - m\frac{ \xi}2 \,\right)
\;,\;\;\;\; m = 0,1,\ldots [2 \pi/3\xi]
\;,
\ee
\be
  M( B_p )
= 2 M \sin\left(\,p\frac{\xi}2 \,\right)
\;,\;\;\;\; p = 1,2,\ldots [\pi/\xi ]
\;.
\ee
We now consider the \sm\ bootstrap to find the \sms\ of the
excited kinks.

\subsubsection{Fusion of the unrestricted \sms\ in the $(1,2)$ gradation}

The equation which describes the
bootstrap procedure in the (1,5) gradation is
\bea
  S_{K_0 K_k}|_{14}
&\propto&
  C_{4|23} \, R_{13}(\e xq^2) \, R_{12}(\e x/q^2) \,C_{23|4}
= \frac{\e xq^2-1}{\e x-q^2}\,R_{14}(-\e x,q)\ ,
\\
  S_{K_k K_0}|_{41}
&\propto&
  C_{4|12}\, R_{13}(\e x/q^2)\, R_{12}(\e xq^2) C_{12|4}
=  \frac{ \e x q^2 - 1}{\e x - q^2}\, R_{41}(-\e x , q)
\label{hkfus12}
\eea
where the indices $1,2,3,4$ indicate in which spaces the matrices act,
$\e = \pm 1$ and $C_{4|23}$ and $C_{23|4}$ are the Clebsh-Gordan
coefficients specifying  the embedding of the triplet representation
space $4$ into the product of the two triplet representation spaces
$2$ and $3$.

In fact, as noted by Smirnov, these relations describe {\em all} the
bootstraps in the $(1,2)$ gradation, and that all higher kink--higher
kink \sms\ have the same group structure,
\be
  S_{K_m\,K_n} \propto R(\, (-1)^{m+n} x\;,\;q)
\;.
\label{eq:skmkn12}
\ee
Since we have already seen that $R(-x,q)$ doesn't have phase
eigenvalues for positive real $x$ and $q$ a phase, we immediately see
that, as for \ato, it is only possible for the unrestricted theory to
have  unitary \sms\ if we must again restrict the coupling constant,
this time to $\xi > 2\pi/3$.

As we have noted, it is not sufficient for the two-particle \sm\ to
have phase eigenvalues, but every $n$--particle \tm\ must
also have phase eigenvalues; this is guaranteed if the two-particle
\sm\ is conjugate to a unitary matrix up to a change of inner products
on the one-particle states, but this is too hard for us to
check. Equally it is impossible for us to check all \tm\
eigenvalues, but we have investigated the three-particle case
numerically and find that for a large set of values of $q$ and
rapidities, the three-particle \tm\ indeed has phase
eigenvalues. Thus we conjecture that this is true for all
multi-particle states and that (to leading order) the finite size
spectrum is real.

\subsubsection{Bootstrap fusion for the ZMS model in the (1,5) gradation}

The bootstrap relation looks a little different in the $(1,5)$ gradation.
If $H$ (\ref{eq:h2}) were to have had only even integer eigenvalues,
then this would make no difference to the bootstrap, but that fact
that it has also odd eigenvalues means that the bootstrap equations
resulting from the two poles $y = \pm q^2$ corresponding to $x = q^4$
are different.

After including the effects of $S_0$, we have two series of poles in
$\theta$ , which we call `even' and `odd', according to
\bea
  y=q^2 && \theta = 2 i \pi/3 - n \xi\;,\;\;\hbox{$n$ odd},
\nn
  y=-q^2 && \theta = 2 i \pi/3 - n \xi\;,\;\;\hbox{$n$ even},
\label{eq:eo}
\eea
Denoting the fundamental kink by $K_0$ and the kinks arising from the
poles in (\ref{eq:eo}) by $K_n$, we see that we have to distinguish
between $S_{K_0 K_{even}}$ and $S_{K_0 K_{odd}}$.
The \Rm\ parts of these \sms\ are now given by
\be
\begin{array}{lclrcl}
y=\!\!\!\!&q^2&:\;\;\;\; &
  A_4^{-1} \, D_{4|23}\, \tR_{13}(\e yq) \,\tR_{12}(\e y/q)\,D_{23|4} A_4
&\!\!=\!\!&
 \frac{ y^2 q^2 - 1 }{y^2 - q^2} \,\tR_{14}(-\e iy)
\;,
\label{hkfus15_1}
\\[3mm]
y =\!\!\!\!& -q^2&:
&
  A_4
  \,\tilde D_{4|23}\,\tR_{13}(i\e yq)
  \,\tR_{12}(-i\e y/q)\,\tilde D_{23|4}A_4^{-1}
&\!\!=\!\!&
  \frac{ y^2 q^2 + 1 }{y^2 + q^2}
  \,\tR_{14}(-\e y)
\;,
\label{hkfus15_2}
\end{array}
\ee
where again $\e=\pm 1$, $D_{23|4}$ etc are Clebsh-Gordan coefficients
for the (1,5) algebra and $A$ is a matrix which rescales the
eigenvectors since $\bP \tR$ no longer degenerates to a pure projector
at $y=\pm q^2$.

Thus we find that
\be
  S_{K_0\,K_{\rm even}} \propto \tilde R(\pm y, q)
 \sim R(y^2,q)
\;,\;\;\;\;
  S_{K_0\,K_{\rm odd}} \propto \tilde R(\pm i y, q)
 \sim R(-y^2,q)
\;,
\label{eq:ss}
\ee
and although there are now four different $\tR$ matrices appearing in
the soliton \sms\,
and e.g.\ $S_{K_0 K_2}$ is no longer proprtional to $S_{K_0 K_0}$, the
eigenvalues of the unrestricted  \sms\ in the (1,5) gradation are the
same as in the (1,2) gradation, and the same comments about the
restriction to $\xi >2\pi/3$ still apply.

\section{Classical scattering in \att}
\label{sec:a22classical}

The classical solitons and breathers in \att\ are not as well
understood as those in \ato. In particular, quantum considerations
suggest that there should be classical breather solutions with all
masses up to twice $M_K$,
the kink mass, but these are hard to find. The
quantum spectrum also suggests that there might be a class of
classical configurations of zero topological charge with masses lying
between $M_K$ and $2 M_K$, but which are somehow to be regarded as
distinct from the breathers, rather behaving like the zero topological
charge component of the kink triplet.
The main difficulty is that there is no consistent way to find
classical solutions of any particular topological charge; secondly,
the \att\ classical solutions can be found as a subset of the \ato\
classical solutions, but with typically twice as many component
`kinks' in the tau functions. One possibility is that the new methods
of Beggs and Johnson \cite{BeggsJ}
may be adapted to find such solutions, but at
present this is an open problem.

As a result it is very hard to compare classical and quantum
scattering in \att\, since so many of the classical configurations are
missing. We have been able to compare the scattering of non-zero
charge solitons and low mass breathers, and there are no discernible
problems -- all the effects of scatterings seem to be able to be
described by a simple time delay, and the resulting time delays agree
with the semi--classical limit of the quantum transition amplitudes
according to the formalism of Faddeev and Korepin \cite{FKor1}.

However, the specific problems we encountered in the quantum \att\
scattering were a result of regradations, which do not affect
transition amplitudes. To see the full effect and to compare these
with classical results, we need to compute the reflection amplitudes
in the semi-classical formalism, something we have not yet done.

We would like to note here that a discussion (we believe new) of how
different gradations can be understood as arising from different
classical actions in the path integral formalism can be found in
appendix \ref{app:regrade}.

\section{
An explicit example of both \ato\ and \att\ Toda theories:
$M_{3,14} + \Phi_{1,5}$
}
\label{sec:m314}

What initially stimulated our interest in this problem was the strange
behaviour of $M_{3,14} + \Phi_{(1,5)}$, and since this is instructive in
many ways, we briefly summarise the situation in this model.
The main point of interest with this model is that there
are a pair of particles $A,\barA$ of mass $M$ and a
particle $B$, or pair of particles $B,\barB$, of mass
$(1 {+} \sqrt{3}) M/\sqrt{2} $, for which the \sm\
$S_{AB}(\theta)$ is not a phase for real $\theta$ \cite{takacs}.

When we try to identify this model with an RSOS restriction of a Toda
theory, we find that it is somewhat special in that conformal field
theory of the `D' modular invariant of the Virasoro minimal model
$M_{3,14}$ is the  minimal model $M\!A_2(3,7)$ of the $W_3$ algebra;
correspondingly the primary field of weight $-5/7$ is both the
$\Phi_{(1,5)}$ primary field of the Virasoro algebra and the
$\Phi_{(11;22)}$ primary field of the $W_3$ algebra. Since perturbing a
Virasoro model by the $\Phi_{(1,5)}$ field  gives the \att\ Toda theory,
and perturbing a $W_3$ minimal model by the $\Phi_{(11;22)}$ field gives
the \ato\ Toda theory, this single model gives an example of RSOS
restrictions of the two theories we have discussed in this paper.

{}From the point of view of \ato\ Toda theory,
this is the RSOS restriction of a model with $\lambda = 4/3$, and the
solitons are entirely eliminated by the RSOS restriction; the
fundamental particles $A$ and $\barA$ of the RSOS restriction
discussed in \cite{takacs,us} are the first breathers $B_1$ and
$\barB_1$.
However the particle(s) $B$ are neither breathers nor
solitons in the generic unrestricted model, and do not appear in the
spectra of Gandenberger or Hollowood. From the point of view of \att\
theory, this is the $(1,5)$ restriction with $\xi {=} \pi/2$;
$A$ and $\barA$
are the only remnants of the fundamental soliton, and
$B$ and $\barB$ are the remnants of the first excited soliton.

This model is instructive in two ways:

Firstly it shows that a restriction of \ato\ to a regime where there
are no kinks does not automatically imply that the theory is unitary.
Even if there is only one RSOS restriction, problems can arise
when one attempts to complete the bootstrap. As we have seen,
there is a particle $B$
arising in the course of the bootstrap whose \sms\ with $A$ are not
phases. The reason why Hollowood and Gandenberger miss this particle is
that in nonunitary theories it is difficult to decide which poles to
invite in the bootstrap. From the point of view of \att\ it was very
natural to conjecture the complete spectrum, and we
found a set of \sm\ amplitudes which is closed
under the bootstrap and is proven to be correct by TCSA and TBA analysis.
The problem of the existence and uniqueness of a closed system of scattering
amplitudes given the \sm\ of some fundamental particle is very nontrivial
and unsolved in general.

Secondly it shows that whether the RSOS restriction of \att\ is
unitary or not depends in part on whether we mean the $(1,2)$ or $(1,5)$
RSOS restrictions -- since, as we discuss in \cite{us}, the $(1,2)$ RSOS
restriction of \att\ at the same value $\xi {=} \pi/2$ gives the
perfectly satisfactory model $M_{6,7} + \Phi_{(1,2)}$ already treated
in \cite{smir}.
We shall treat this and other problems connected with RSOS
restrictions in \cite{us2}.

\section{Conclusion}
\label{conclusions}

First we review the results for \ato\ and \att\ Toda theories
separately; we then consider their implications and discuss whether
one can find a coherent picture for Toda theories into which they fit.

\subsection{\ato\ Toda theory}
\label{a21conclusions}

We first investigated the unrestricted \sms\ of the fundamental and
excited solitons in both the homogenous and spin gradations; we found
that the \sms\ describing the scattering of solitons of the same
species had phase eigenvalues, and hence are unitary with an
appropriate inner product, but that this cannot simply be achieved by
changing the inner products of the one particle states.
This leads to non-phase eigenvalues for the three-particle \tm\ and a
complex finite size spectrum.
Even more clearly,  the \sms\ describing the
scattering of conjugate species did not have phase eigenvalues and
hence cannot be made unitary by any change of inner product.
The only known way to obtain a unitary theory with both solitons and
antisolitons is by quantum group restriction to an RSOS theory.

We next investigated the scattering of solitons and breathers and
again found that the amplitudes were not phases for real rapidity
differences. The difference now is that since breathers are scalars
under the quantum group, these \sms\ are unaltered by RSOS
restriction.

However, de Vega and Fateev \cite{VFat} have presented a set of \sms\
which  they claim correspond to unitary theories, including
$a_2^{(1)}$ Toda theory in particular. The answer lies in the fact
that their \sms\ correspond to particular values of the coupling
constant for which, as Hollowood pointed out \cite{Holl1}, it
is possible to remove the breathers consistently.
According to Hollowood, there are no poles corresponding to breathers
in the physical strip provided
\be
  0 \leq 3 \lambda \leq 1
\;.
\ee
The \sms\ of de Vega and Fateev correspond to perturbations
of the unitary minimal models of the $W_3$ algebra with
\be
  \frac{\beta^2}{4 \pi} =  \frac{m}{m+1}
\;,
\ee
and as a result, the unitarity-violating breather-soliton \sm\ element
does not occur as the breathers are entirely decoupled.

If $\beta^2/(4\pi) < 3/4$ then breathers do arise as
poles in the physical strip and cannot be ignored,
and in this case the only possibility is to remove the entire set of
solitons from the theory by choosing the values of $\beta$
appropriately.  Whenever $q^3 = \pm 1$, the RSOS restriction ensures
that there is only a single topological charge possible, and hence the
kinks are entirely removed from the spectrum. We see that this
corresponds to
\be
  \frac{\beta^2}{4\pi} = \frac{3}{q}
\;,
\ee
i.e. corresponding to the non-unitary minimal models $M_{3,q}$ of the
$W_3$ algebra.
However, we have already seen in section \ref{sec:m314} that this does
not automatically ensure that the spectrum is real.

For any other value of $\beta$, the breather arises as a bound state
from a pole in the physical strip, and the RSOS procedure, while
changing restricting the soliton vacua to a finite set and changing
the \sms\ of the kinks, cannot affect the kink-breather \sm\ as the
breather is a quantum group singlet. As a result, we expect that all
these theories will have a complex spectrum.

Since this is such a strange result, that all quantum soliton--breather
scatterings are nonunitary, we also investigated the classical
theory. We found that the classical results exactly mirror the quantum
results, in that the time delay (the analogue of the \sm) is also
complex in soliton--breather scatterings. We also found that the
soliton--breather solutions are frequently singular, and that it does
not appear possible to construct a consistent theory of breathers and
solitons which will lead to purely regular solutions.

We believe that this is a general property of affine Toda theories
which contain non-self-conjugate particles, and is related to the
classical scattering properties of breathers and solitons, which we
describe in section \ref{sec:qcl}.

(The soliton--soliton scattering in the classical theory is slightly
problematic in that there are solutions with isolated singularities,
but the time delays are real and the semi-classical limits of the
quantum \sms\ are unitary.)

\subsection{\att\ Toda theory}
\label{a22conclusions}

First of all we investigated the unrestricted \sms\ for the
fundamental soliton and the excited solitons in the homogeneous and spin
gradations. We found that the \sms\ of the fundamental solitons had phase
eigenvalues, but the \sms\ of the fundamental soliton-first excited
soliton never had phase eigenvalues.

This means that for an unrestricted theory to have a real finite size
spectrum, one has to go to a repulsive regime, in which there are no
excited solitons in the spectrum.

It is possible that RSOS restriction can cure the problems in
soliton--excited soliton scattering, but as shown clearly in section
\ref{sec:m314}, the details will depend on the gradation, and we shall
turn our attention to these and other issues of RSOS theories in
\cite{us2}.

We found no evidence of any problems in the classical theory, apart
from the absence of certain solutions one might expect to exist from
quantum considerations -- a problem we return to in section
\ref{sec:qcl}.

\subsection{Quantum vs.\ classical Toda theory}
\label{sec:qcl}

There is a general issue, which is to what extent classical and
quantum Toda theories with imaginary couplings are sensible physical
theories. While it was recognised that the unrestricted theories were
apparently non-Hermitian, it was thought that these problems could be
cured by restriction -- to a suitable space of classical solutions in
one case, and by a quantum group restriction in the other -- and that
the resulting models would be a good description of perturbed
conformal field theories.

All the models which had been investigated prior to \cite{us} had been
found to have an entirely real spectrum, as they did indeed correspond
to unitary \sms\ found by one method or another.
However, as we found in \cite{us}, this is not necessarily the case --
as indeed one should expect. There is no reason why the formal
Hamiltonian of a perturbation of a non-unitary conformal field theory
should have a real spectrum, even if the theory appears integrable.

What is most remarkable about the result of \cite{us} is that the
spectrum of the perturbed Virasoro minimal model $M_{3,14} +
\Phi_{(1,5)}$ is in excellent agreement with the nonunitary \sms\
obtained by RSOS restriction and the bootstrap, even
including the complex eigenvalues of the Hamiltonian which result from
\sm\ elements which are not phases.
Furthermore, the complex \sms\ describing soliton--breather scattering
in \ato\ are in perfect agreement with the (complex) classical time
delays, even though the corresponding classical solutions may be
singular.

Thus we can conclude that the integrable structure and quantum group
symmetry which is generically present in affine Toda theories and in
perturbed conformal field theories is still present even when the
spectrum is no longer real, and that a formal \sm\ description still
remains valid even when (\ref{eq:usab0}) or (\ref{eq:usab1}) no longer
hold.

As noted by Gandenberger and MacKay in \cite{GMac}, there is a close
correspondence between the transmission coefficients $X_{ab}(k)$ which
play an important role in the semi-classical method, and the
soliton-$a$--breather-$b$ \sm\ $S_{ab}$, namely that
\be
  X_{ab}( m_a \sinh(\theta) ) = \lim_{\beta \to 0} S_{ab}(\theta)
\;.
\ee
This can be easily checked for the case of $a_2^{(1)}$, as the
transmission coefficients for $a_n^{(1)}$ are \cite{Holl2}
\be
  X_{ab}(k)
= \frac{ i k  - m_a \cos((a-b)\pi/(n+1) ) }{ i k  - m_a \cos((a+b)\pi/(n+1) )}
\;,\;\; m_a = 2 m \cos(a \pi/(n+1))
\;.
\ee
For $a_2^{(1)}$, we find as required that
\be
  X_{11}(k) = \frac{ \sinh\theta - i }{\sinh\theta -i/2}
= \lim_{B \to 0} S_{AB_1}(\theta)
\;.
\ee
The transmission coefficients for all simply-laced and twisted affine
Toda theories have been found in \cite{OTUn1,KOli2}, and they do not
satisfy (\ref{eq:usab1}) exactly when both species are
not-self-conjugate; such particles occur for $a_n^{(1)}$,
$d_{2n+1}^{(1)}$ and $e_6^{(1)}$ theories. Since this property will be
inherited by the soliton--breather \sms, we expect that in all these
cases there is no unitary \sm\ possible, even after RSOS restriction,
whenever both the relevant solitons and particles remain in the
spectrum.

There is also the problem of missing solutions: it is well known that
for a general affine Toda theory the simplest ansatz for soliton
solutions do not fill out the representations and there some
topological charges missing, but we also mention that it is possible
that this problem also applies to breather solutions and that not all
the classical solutions expected on quantum grounds are known at
present. It is possible that this will be cured by the new methods of
Beggs and Johnson \cite{BeggsJ}.

It is indeed interesting that the unrestricted \sms\ of the
fundamental soliton--soliton scattering in both
\att\ and \ato\ in the repulsive regimes have only phase eigenvalues,
and we reinvestigate RSOS theories in this light in \cite{us2}.

Finally, we should like to note that \sms\ do not only appear in
scattering theories and statistical field theory, but other physical
applications such as diffusion--annihilation problems (see e.g.\
\cite{Henkel}) and it may well be that a non--real spectrum is even
necessary in such applications.

\section{Acknowledgments}

GMTW would like to thank
H.W. Braden, E. Corrigan, G. Delius, P.E. Dorey, E.B.\ Davies,
M. Freeman, G. Gandenberger, U. Harder, P.R. Johnson, H.G. Kausch,
N.J. MacKay and R. Sasaki for enlightening discussions and INFN
Bologna for hospitality and support from INFN (Italy) grant
\emph{Iniziativa Specifica} TO12.
We would also like to thank F. Ravanini for discussions,  SISSA for
hospitality during several stages of this work, and the EU TMR network
FMRX-CT96-0012 for support, and especially the organisers of the
Miramare98 meeting where this work was completed.

GMTW thanks EPSRC for an advanced fellowship and support under grant
GR/K30667.
GT thanks INFN for a postdoctoral fellowship, and Hungarian grants
FKFP 0125/1997 and OTKA T016251 for partial support.

\vspace{1cm}
\appendix

\section{Analysis of singularities}
\label{app:sings}

To determine for which parameters \ato\ and \att\ breathers and
breathing kinks are singular, it is necessary to decide for which
values of $\delta, u, X \in R$ the following equation
\be
  1 + \exp( i \delta)\left( z u + z^*/u \right)
 + \exp(2 i \delta) X z z^* = 0
\;,
\label{eq:s1}
\ee
has a solution for some value of $z \in C$.
We first turn this complex equation into two real equations for the
real variables $x,y$ where $z = x + i y$. These two equations are
\bea
0 &=&  1 + x\,\cos\delta \,( u  + 1 / u)
    + y\,\sin\delta\,( - u + 1 / u) + X (x^2 + y^2) \cos(2\delta)
\label{eq:s2}
\\
0 &=& y\,\cos\delta( u - 1/u) + x\,\sin\delta\,(u+1/u)
    + X (x^2 + y^2) \sin(2\delta)
\label{eq:s3}
\eea
These are the equations of two circles in the plane, with radii
$r_1,r_2$ and distance $d$ between the centres given by
\be
\begin{array}{rcl}
 (r_1)^2 &=& \ds \frac{ \tau + (2 - 4 X) X }{4 X^2 c^2 }
\\[3mm]
 (r_2)^2 &=& \ds\frac{ \tau - 2 c}{4 X^2 s^2}
\\[3mm]
 d^2   &=& \ds\frac{ \tau - 2 c}{4 X^2 s^2 c^2}
\end{array}
\ee
where
\[
  \tau = u^2 + 1/u^2
\;,\;\; c = \cos(2 \delta)
\;.
\]
For eq (\ref{eq:s1}) to have a solution we need these two circles to
have real solutions and to intersect, that is
\be
\begin{array}{c}
 (r_1)^2 > 0 \;\;\;\;\;\;\;\;  (r_2)^2 > 0
\\[2mm]
 (r_1^2 - 2 r_1r_2 + r_2^2 - d^2)
 (r_1^2 + 2 r_1r_2 + r_2^2 - d^2) < 0
\end{array}
\ee
These three equations, after much manipulation, can be put in the form
\be
\begin{array}{c}
  \tau > (4X - 2)c \;\;\;\;\;\;\;\;
  \tau > 2 c
\\
  ( \tau - (2X(1+c)-2) )( \tau - (2X(c-1)+2)) > 0
\end{array}
\ee
which, together with the obvious equation
\be
  \tau \geq 2
\ee
are the full set of equations giving the limits on $\tau$ in terms of
$\delta$.

\section{Regrading in quantum and classical pictures}
\label{app:regrade}

Let us finally comment on some properties of the regrading
transformations in the context of quantised affine Toda theories.

In a general affine Toda theory, the minima of the potential are
labelled by weights $\lambda_a$ of some Lie algebra (which for
untwisted affine Toda theories is the dual of the corresponding finite
dimensional algebra), and so typical \sm\ elements are labelled.
\be
  \kS {},{},\lambda_a,\lambda_b,\lambda_c,\lambda_d.
\;.
\ee
The consequence of a changing the gradation $\gamma$ in eqn.\
(\ref{eq:regrade}) is to change this \sm\ element by a factor
\be
  \exp(\, \theta \,
       \delta\gamma\cdot(\lambda_a + \lambda_d - \lambda_b - \lambda_c)
       \,)
\;.
\label{eq:regrade2}
\ee
This can be thought of as a change in the action $I$, and such a
term will change the scattering amplitude by a factor
$\exp( \, i \, \delta I\,)$. The correct form of $\delta I$
is
\be
  \delta I
= \int {\rm d}^2x \,\sqrt{-g\,}\,R\,(\delta\gamma\cdd\Phi)/2
\;,
\label{eq:regrade3}
\ee
where the integral is over all spacetime, $g$ is the metric, $R$
the scalar curvature and $\Phi$ the Toda field appropriately normalised.
This is zero in the interior of space time, but Minkowski space has
singularities at infinity and this integral can give a finite answer
by the following prescription.

Consider the following variables $\rho,\theta,\phi$ given in terms of
$x$ and $t$ by
\be
 t = \rho \cosh\phi
\;,\;\;
 x = \rho \sinh\phi
\;,\;\;\;\;
 \tanh{\theta/2} = 1/\rho
\;.\;\;\;\;
\ee
These variables do not cover the whole of the forward light-cone, but
a subset which does, however, include the whole of the forward
time-like infinity, which is the line $\theta = 0,
-\infty<\phi<\infty$, and for which the metric is
\be
  \d s^2
= \d\rho^2 - \rho^2\d\phi^2
=  \d s^2 = \Omega^2 \left( \d\theta^2 - \sinh^2\theta\d\phi^2\right)
\;,\;\;
  \Omega = \rho /  \sinh\theta
\;.\;\;
\ee
In these variables we can calculate the integral (\ref{eq:regrade3})
using a test function, using the result
\be
   \sqrt{ - g} R
 = - 2
   \frac{\partial}{\partial \theta}\left(
   \sinh\theta
   \frac{\partial}{\partial \theta}
   \log\left( \Omega \sinh\theta \right)\right)
\;,
\ee
so that
\be
\;\int_{-\infty}^\infty \!\!\d\phi
  \int_{0}^\infty \!\!\d\theta
  \;\;\sqrt{-g} R \;\;f(\phi)
=
2 \int_{-\infty}^\infty \!\!\d\phi
  \;\; f(\phi, \theta = 0)
=
2 \int_{-\infty}^\infty \!\!\d\phi
  \;\; f(\phi, t = \infty)
\;.
\ee
There is a similar contribution from backward time-like infinity which
gives us the total contribution from the two time-like infinities
to the additional terms in our action $\delta S$
\begin{equation}
\delta I =
\int \d^2x\, \sqrt{-g} R \,(\delta\gamma\cdd\Phi)/2 =
  \int_{-\infty}^\infty \!\!\d\phi
  \;\; \delta\gamma\cdd
     \left(\;\Phi(\phi, t = \infty) - \Phi(\phi, t=-\infty)\;\right)
\;.
\label{eq:1}
\end{equation}
To evaluate this for a two-particle scattering amplitude as in figure
\ref{fig:2ps}, we first note that
a particle of rapidity $\phi_0$ intersects time-like infinity at
$\phi = \phi_0$, and evaluating (\ref{eq:1}) with a suitable cutoff to
regulate the $\phi$ integral gives (the cutoff disappears as we would like)
\be
  \delta I
= (\theta_1 - \theta_2) \;
  \delta\gamma\cdd( \lambda_a + \lambda_c - \lambda_b - \lambda_d )
\;,
\ee
which is exactly the result of a regradation in the quantum group picture.

{
\begin{figure}[ht]
\[
\epsfxsize=5truecm
\epsfbox{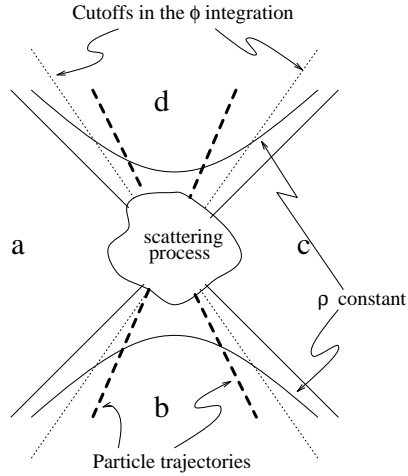}
\]
\caption{\small
A two--particle scattering showing the cutoffs in $\phi$ required to
give a finite answer for $\delta S$}
\label{fig:2ps}
\end{figure}
}

%%%%%%%%%%%%%%%%%%% BIBLIOGRAPHY %%%%%%%%%%%%%%%%%%%%%%

  \end{document}